\UseRawInputEncoding
\documentclass{article}

\usepackage{lipsum}

\usepackage{microtype}
\usepackage{graphicx}
\usepackage{subfigure}
\usepackage{booktabs} 
\usepackage{hyperref}

\usepackage{tabularx}
\usepackage{paralist}
\usepackage{enumitem} 
\setlist{nosep,leftmargin=*}

\usepackage[accepted]{icml2025} 

\usepackage{amsmath, amssymb, amsthm, mathtools}
\usepackage[capitalize,noabbrev]{cleveref}

\usepackage{xcolor}
\newcommand{\dnote}[1]{\textcolor{blue}{$\ll$\textsf{#1 | Dafna}$\gg$}}
\newcommand{\danote}[1]{\textcolor{brown}{$\ll$\textsf{#1 | Dan}$\gg$}}
\renewcommand{\dnote}[1]{}
\renewcommand{\danote}[1]{}

\newcommand{\remove}[1]{}

\usepackage{enumitem}
\setlist[itemize]{noitemsep} 
\theoremstyle{plain}

\theoremstyle{definition}

\theoremstyle{remark}

\icmlkeywords{Machine Learning,UK Biobank, Medical, Language models, Literature based discovery, creativity, novelty, hypothesis generation, Bioinformatics, Disease, SHAP, feature selection}
\usepackage{listings}
\title{InterFeat: A Pipeline for Finding Interesting Scientific Features}
\icmltitlerunning{InterFeat: A Pipeline for Finding Interesting Scientific Features}

\author{
  Dan Ofer\thanks{The Hebrew University of Jerusalem, Israel. \texttt{dan.ofer[at]mail.huji.ac.il}} \\
  \and
  Michal Linial\thanks{Department of Biological Chemistry, Institute of Life Sciences, The Hebrew University of Jerusalem, Israel. \texttt{michal.linial@mail.huji.ac.il}} \\
  \and
  Dafna Shahaf\thanks{Department of Computer Science, The Hebrew University of Jerusalem, Israel. \texttt{dshahaf@cs.huji.ac.il }}
}
\begin{document}

\maketitle
\begin{abstract}
Finding \emph{interesting} phenomena is the core of scientific discovery, but is a manual, ill-defined concept. We present InterFeat, an integrative pipeline for automating the discovery and ranking of \textbf{inter}esting \textbf{feat}ures (\textbf{InterFeat}) in structured biomedical data. The pipeline combines machine learning, knowledge graphs, literature search and Large Language Models.
We formalize ``interestingness" as a combination of novelty, utility and plausibility.
In a 2011 time-split, InterFeat surfaced known risk factors years before the literature: across eight major diseases, up to 21\% of utility‑-filtered candidates appeared in knowledge bases only after the cut‑off.
In human evaluation, four senior physicians independently annotated 118 candidates for gallstones, esophageal cancer, heart attacks and gout for novelty, plausibility, utility and overall interestingness. 28\% were deemed interesting. For highly-ranked candidates, 40--53\% were interesting, vs 0--7\% for a SHAP baseline.
InterFeat addresses the challenge of operationalizing ``interestingness" scalably for any target with existing literature. Code and data: 
\href{https://github.com/LinialLab/InterFeat}{https://github.com/LinialLab/InterFeat}  
\end{abstract}



\section{Introduction}
\label{sec:introduction}

Finding interesting phenomena in data is the essence of discovery. Yet the notion of “interestingness” is surprisingly elusive, requiring subjective human judgment and lacking the relatively well-accepted metrics that concepts such as “statistical significance” enjoy. 

We build a pipeline that extracts interesting hypotheses about connections between features and target diseases, including the direction of effect and potential underlying mechanisms. 
We identify three core concepts that lie at the heart of interestingness: novelty, utility (usefulness), and plausibility (the existence of an underlying explanatory mechanism). 


The exponential growth of data and literature has not been accompanied by a corresponding growth in insights, and finding interesting, actionable insights from data remains a challenging task.
Many now-obvious discoveries, such as the link between contaminated water and disease or handwashing, were overlooked for millennia. Hand hygiene gained acceptance only after germ theory, and H. pylori as the cause of ulcers was ridiculed until \citet{marshall_attempt_1985}'s self-experimentation. Lithium, now essential for treating bipolar disorder
would sound absurd if proposed naively \cite{cade_cades_2007}. These insights existed in the data but were missed or dismissed due to innate biases, insufficient explanatory frameworks or statistical rigor.

This work presents an integrative framework for quantifying and automating the discovery of interesting features in scientific datasets. We focus on identifying disease risk factors from the biomedical UK BioBank (UKB), although the underlying principles and methodologies are generalizable to other populations and non-medical domains. 
Our contributions are:

\begin{compactitem}
\item We combine machine learning and natural language processing to create an expressive and easy to use pipeline ("InterFeat") for finding interesting features.
\item The InterFeat pipeline leverages structured data from electronic health records, biomedical ontologies and Knowledge Graphs (\textbf{KG}), scientific literature and Large Language Models (\textbf{LLM}s) to systematically identify, rank and explain features with high potential for discovery.
\item We ground our approach in a formal definition of interestingness, integrating statistical and literature-based discovery approaches with LLMs to flexibly assess and rank features based on novelty, plausibility, and utility criteria, in relation to a target.

\item We leverage LLMs to generate explanatory
rationales and mechanistic explanations for candidate hypotheses
, guiding researchers' prioritization. 

\begin{figure*}[t!] 
    \begin{minipage}[htb!]{.4\textwidth}
  \includegraphics[width=\linewidth]{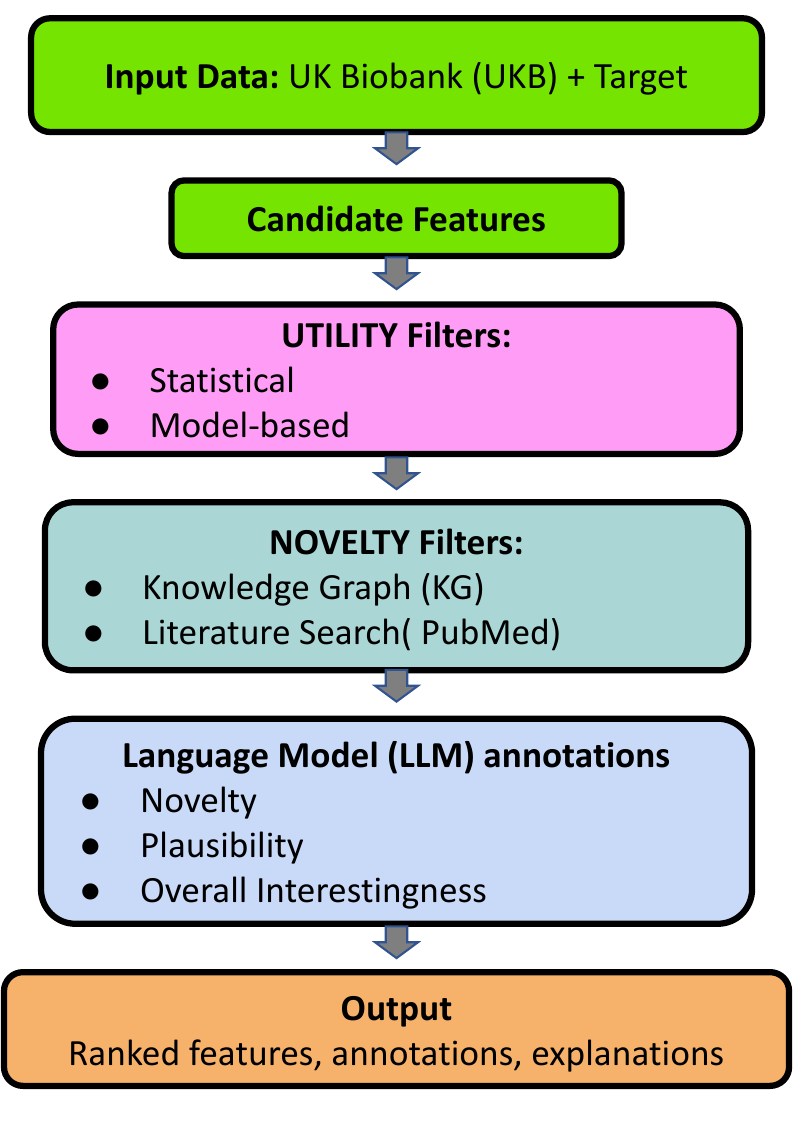}
  \label{fig:sub1}
\end{minipage}%
\ \ 
\begin{minipage}[htb!]{.55\textwidth}
\footnotesize
\fbox{\parbox{\linewidth}{
\textbf{Input hypothesis:} \\[0.3em]
Feature: Vitamin D, Target Disease: Oesophagus Cancer,
Correlation: Negative \\[-0.2em]
\hrule
\vspace{0.45em}

\textbf{Explanation:} \\[-0.45em]

\textbf{1. Novelty:} Association not well-documented in literature. While Vitamin D has been studied with other cancers, oesophagus cancer is not mentioned. 
\smallskip

\textbf{2. Plausibility:} The biological plausibility of Vitamin D affecting oesophagus cancer risk is not strongly supported by current scientific understanding. Vitamin D is known to play roles in cell growth and immune function, which could theoretically influence cancer risk. However, without specific mechanisms or pathways identified for oesophagus cancer, the plausibility remains weak. The negative correlation observed does not align with typical Vitamin D-cancer risk patterns seen in other cancers, adding to the uncertainty. 
\smallskip

\textbf{Summary:} Given the novelty of the association and the potential for future research to uncover new insights, the feature is somewhat interesting to researchers. However, the lack of plausibility and utility tempers the level of interest. \\[-0.2em]

\hrule
\vspace{0.5em}
\textbf{Output Annotations:} \\[0.2em]
Novel: 3, Plausible: 2, Interesting: 3, 
Explanation:".."
}} 
  \label{fig:sub2}
\end{minipage}
    \caption{{\bf Left: InterFeat pipeline.}  (i) Target and features are extracted from a dataset (Here, the UK Biobank). (ii) Statistical and model-based methods are combined to retain features with predictive value (utility). (iii) UMLS-CUI linked entities are extracted and linked to a knowledge graph, to exclude known associations. Literature mining, via PubMed, removes frequent co-associations. Finally, (iv) language models (optionally augmented with relevant, retrieved texts) annotate the remaining features for novelty, plausibility 
    and overall interestingness. Outputs include a ranked list of features with annotations and natural language explanations. {\bf Right: LLM Annotation Example.} The input consists of a candidate feature-disease association. The LLM provides separate judgements (combined here for clarity) for novelty, plausibility, and overall interest, scored 1-4. This specific feature was confirmed as interesting, novel and useful by experts.
    Text edited for clarity.}
    \label{fig:Pipeline}
\end{figure*}

\item We release our code and a novel expert-labeled multi-disease dataset of interesting biomedical features, with explanations and human validation.

\item 
Using the UK Biobank (370,000+ patients), we demonstrate InterFeat’s effectiveness in uncovering previously undocumented risk factors across 8 major diseases, highlighting its potential to accelerate scientific discovery 

\item In temporal validation using a 2011 cutoff, over 8 diseases, up to 21\% of features retained after pipeline utility filtering were not reported in the literature until after 2011, demonstrating InterFeat’s ability to surface novel insights years before their documentation 
\item InterFeat surfaced features experts rated as interesting: across 118 candidates and 4 diverse, major diseases, 28\% were judged interesting by physicians, and among top-ranked hypotheses InterFeat achieved 40-53\% vs. 0-7\% for a SHAP-only baseline.
\end{compactitem}


\section{Related Work}
\label{sec:related}

\textbf{Automated Hypothesis generation} aims to systematize the traditionally intuitive process of discovery \cite{tong_automating_2024,spangler_automated_2014}. Methodologies, such as literature-based discovery (LBD), aim to identify missed connections between concepts and findings, thereby uncovering novel hypotheses \cite{henry_literature_2017,swanson_fish_1986,Voytek2012b}. However, traditional hypothesis generation approaches face several limitations: i) Directionality ignorance: These methods often treat associations between concepts as bidirectional, ignoring the direction of effect. For example, smoking reducing the risk of a disease would be novel, useful and interesting, while the inverse would not. 
ii) Ontology dependency: Many approaches rely on a standardized ontology and linkage to define co-occurrence, which limits it in terms of the source ontology and precision of linkage \cite{Voytek2012b}. Recent studies have begun to address these limitations using deep learning and graph methods to improve the flexibility of LBD approaches \cite{wang_scimon_2024,moreau_literature-based_2021}.

Deep learning-based large language models have been used to automatically generate ideas and hypotheses and can flexibly capture unstructured relationships \cite{zhou_hypothesis_2024,sybrandt_are_2018,tong_automating_2024,qi_large_2024,spangler_automated_2014,wang_scimon_2024}. They have been shown to have near expert level scientific and medical understanding in some tasks, albeit when relying on known knowledge (e.g. differential diagnosis) \cite{brodeur_superhuman_2024,matsumoto_kragen_2024,lievin_can_2024,shringarpure_large_2024,qi_large_2024,ofer_automated_2024}. However, their tendency to hallucinate unfeasible, or nonsensical ideas makes them insufficiently reliable to be a “Great Automatic Grammatizator” \cite{dahl_great_1997} for ideas without manual validation \cite{farquhar_detecting_2024}. The use of actual features as a “starting point” may reduce hallucinations, due to the more limited hypothesis space \cite{wu_medical_2024,bechard_reducing_2024}. 

In practice, the starting point of many researchers looking for interesting connections in their data is statistical machine learning methodologies such as {\bf  feature selection} \cite{jeong_llm-select_2024,maor_system_2019,Domingos2012,madakkatel_combining_2021}. Feature selection approaches focus on predictive power or statistical significance \cite{guyon_introduction_2003,Breiman1999,blum_selection_1997}. This includes the life sciences, such as predicting mortality, Endometriosis, scientific trends, Depression, Heart attacks and viral-proteins \cite{blass_revisiting_2022,cohen_icu_2021,bilu_predicting_2023,moore_xgboost_2022,michael-pitschaze_detecting_2024,alaa_cardiovascular_2019, ofer_whats_2024}. There are many works using machine learning, and SHAPley values \cite{lundberg_unified_2017} have been applied to the UK Biobank to find risk factors, but most approaches rely on manual analysis of candidates, typically from a list of features sorted by model importance \cite{blass_revisiting_2022,allwright_ranking_2023,allwright_machine_2023,allwright_machine_2023,lugner_identifying_2024,madakkatel_combining_2021,lugner_identifying_2024}. LLM-Select \cite{jeong_llm-select_2024} used LLMs to select features by description and task, but again, only for predictive power.

\section{Problem Definition}
\label{sec:problem_definition}

Given a set of datasets over the same set of biomedical features and a target feature $y$, our goal is output a ranked list of interesting simple hypotheses of the form ``$x$ is related to $y$, with a negative/positive correlation'', together with potential mechanisms underlying the hypothesis. For example, in the case of medical data, the target $y$ often represents whether a patient will develop a specific disease. Features $x$ are structured patient-level variables from the data, such as age, biomarkers (Vitamin D levels), genetic risk scores, questionnaires (smoking), medical history (age of asthma diagnosis, medications), etc.

To formulate a notion of interestingness, we are inspired by creativity literature, which frequently conceptualizes innovation as a confluence of {\bf novelty} and {\bf utility}  \cite{glover2013handbook, amabile2018creativity,silberschatz_what_1996}. In other words, a creation is deemed innovative if it is both original and valuable or useful. 
Similarly, we define ``interesting'' hypotheses in scientific data if they satisfy the following criteria:

\begin{compactenum}
    \item \textbf{Novelty:}  $x$ should not be \emph{established} in the literature or canonical knowledge bases as linked to the target $y$. Alternatively, a hypothesis might be considered novel if there is a known connection between $x$ and $y$ but the direction of effect implied by the hypothesis is controversial or unestablished. A feature which is closely related to another known association may not be considered novel (e.g., cigarette vs. cigar smoke). 
    \item \textbf{Utility:} $x$ must have predictive power, adding useful information for predicting $y$. Note that in some use cases ``utility'' also implies that $x$ needs to be actionable (e.g., smoking affects the risks of many diseases, and it can be changed). 
        \item \textbf{Plausibility:} 
        In addition to the criteria inspired by creativity research, in science we believe another critical criterion is plausibility -- $x$ is consistent with current knowledge, and has a theoretical explanation. 
 Medical data is particularly rife with spurious correlations, many of which are spurious or simply reflect underlying confounding factors. Thus, researchers tend to prioritize investigating correlations with plausible mechanisms.        
\end{compactenum} 

We operationalize these requirements by formalizing notions of novelty and utility, integrating well-known metrics (e.g., mutual information) with additional LLM input. LLMs also suggest mechanisms and explanations for each score.
While we are not the first to combine notions of novelty and utility, we propose an integrative, configurable approach that we hope will be adapted by practitioners and serve as a vehicle for new scientific discoveries.

\section{Methods}
Our pipeline is summarized in Figure \ref{fig:Pipeline}. Here, we provide implementation details. Code and annotated datasets are available:
\href{https://github.com/LinialLab/InterFeat}{https://github.com/LinialLab/InterFeat}.
UKBB or SemMed raw data are unavailable due to licensing.

Importantly, there are various ways to formulate novelty and utility. Our pipeline brings together the most prominent formulations, providing an intuitive way to configure and select those best suited for specific use cases. 

\subsection{Data: UK Biobank}

We use the UK Biobank health records dataset as our main structured data source \cite{sudlow_uk_2015}. The dataset contains $\sim$1681 patient covariates (medical record history,  diagnostic results, medications, socioeconomic variables, genomic factors, lifestyle, etc') measured at the time of each patient's initial intake (2009--2011), with ICD-10 medical diagnoses recorded through 2022, for 370K adult patients. ICD-10/ICD-10-CM codes were also mapped to their phenotypes/Phecodes as additional covariates.  

\subsection{Extracting Candidate Features}

    We clean and encode the raw UKB data into a structured format with $\sim$3721 features.
     Features with missing values were mean-imputed, and a "missing" feature flag was added. Features without at least 30 non-missing values are dropped.
    Optionally, our pipeline removes redundant features using correlation feature selection. In interpretability use-cases, it is common to remove highly correlated features to reduce redundancy. A popular default is 0.8-0.95 for the Pearson correlation coefficient \cite{hall_correlation-based_1999,Domingos2012}. We use a 0.9 threshold, so that features with strong linear relationships are dropped as redundant, using the feature-engine library of \citet{galli_feature-engine_2021}.

\subsection{Utility Filter}

The pipeline predicts whether a patient will be diagnosed in the future with a given disease (specified by ICD-10 medical codes).
To help mitigate confounding by age, sex, and BMI, we optionally apply Inverse Propensity Weighting (IPW) on the negative samples \cite{chesnaye_introduction_2021}. The predicted probabilities are used as sampling weights for IPW, and the negatives are resampled down to a given ratio (9:1)
(\ref{tab:PipelineStats}).

We allow users to choose between several utility filters to remove features with no predictive strength for \( y \), each with a corresponding threshold. Specifically,

\begin{compactitem}
\item $p$-value under a univariate test: \\
$\mathrm{pVal}(x,y) \;\le\; \theta_{p}$
\item Mutual information between $x$ and $y$: 
\\ 
$\mathrm{MI}(f,y) \ge \theta_{\mathrm{MI}} $
\item Model-based feature-importance score (e.g., global SHAP): 
$\mathrm{FImp}(f,y) \ge \theta_{\mathrm{FImp}}$
\end{compactitem}

MI and FImp can ascertain non-linear effects. FImp reflects whether a feature is used by a trained predictive model(s), e.g., a boosting tree, unlike p-value.
Users can choose criteria, thresholds and also whether to treat them as a conjunction (all) or disjunction (any). 
After some exploration, we chose lenient thresholds for our experiments: \( p\text{-value} < 0.2 \), \( \text{MI} \geq 10^{-3} \), or \( \mathrm{FImp} \geq 10^{-4} \). In our selected configuration, a feature $x$ passes the utility filter if it met any of the three criteria. 
\subsection{Novelty Filter}
\label{sec:novelty}
Our pipeline supports two ways to filter for novelty, both based on scientific literature. 
\paragraph{KG-based Filter.}
We link features and target diseases to UMLS Concept Unique Identifiers (CUIs) \cite{bodenreider_unified_2004} using scispaCy \cite{neumann_scispacy_2019} and edges in SemMedDB v43 \cite{kilicoglu_semmeddb_2012}, a KG of 130 million semantic predications (subject–predicate–object triples) from 37 million PubMed citations. 
Features and targets are represented by sets of linked entities (\(E(x)\) and \(E(y)\)), extracted using named entity recognition and linkage to entities in the KG (here, UMLS CUIs). If a feature $x$ is already directly connected to the disease $y$ in the KG (with sufficient evidence), we mark it as “known” (i.e., \emph{not novel}) and exclude it. To reduce the chance of false predicates, 
we filtered SemMedDB for predicates that had at least 2 unique citations as evidence, leaving 12.9 million. We treat the graph as unidirectional and ignore the type of predicate. scispaCy's (V5.5) “en\_core\_sci\_lg” entity recognition model was used, with a 0.88 threshold and 3 max entities per candidate, following recommendations for high-precision biomedical entity linking \cite{neumann_scispacy_2019,soldaini_quickumls_2016}. 
A predefined list of irrelevant high level categories are excluded by regex (e.g. "Qualification", “Disease”, "Unit").

Domain-specific semantic similarity is computed between each feature and candidate entities, using a pretrained biomedical sentence-level language model (Biolord \cite{remy_biolord_2022}), measured as cosine similarity. This is used to further remove candidate entities with very low (defined as 0.1$<$) similarity to the feature, and later to define ‘strongly linked’ entities (e.g., ‘alcohol’ and ‘alcoholism’). 
Features were filtered out if all their linked entities were directly linked (1-hop) to the target(s) in the KG, or if they had at least one strongly linked entity ($\geq\theta_{\textit{sim}}$ similarity) with a direct connection to the target(s). 
In our experiments, we chose a threshold of $\theta_{\textit{sim}}=0.4$ as "strongly linked".
\paragraph{Literature-based Filter:}
Text mining is used to ascertain if the co-occurrence of features and disease is already established in the literature. This reflects the typical human search process: “Are there already papers about $x$ and $y$?”.

PubMed is a large literature database of over 37 million published scientific and specifically biomedical works (as of 10/2024). We query the PubMed search API (including automatic term expansions) for publication counts of each feature, the target, and their co-occurrence ($x$ AND $y$). If the pair is co-mentioned less than an absolute threshold $\theta_{\text{lit}}$ or less frequently than expected by random chance (via one-way Fisher’s Exact Test, \textit{p}$<\theta_{\text{pval}}$), it is retained. 
Features with less than 20 hits in the database are left unfiltered (these could include, for example, recently coined terms).
Again, after experimentation, we chose to use relaxed default threshold: \(\theta_{\text{lit}} = 4\), \(\theta_{\text{pval}} = 0.4\). 

\noindent {\bf A note on thresholds. }
In both novelty filters, we prefer high recall, filtering out only clearly non-novel features while retaining borderline cases. This prioritizes precision in exclusion, minimizing the risk of discarding under-explored but potentially meaningful findings.

\subsection{LLM Annotations}
\label{sec:llm-annotations}
To refine and rank filtered features, we use LLMs as an extra layer of information. Due to the nature of language models, we chose to focus on novelty and plausibility: language models are very effective for processing and internalizing vast amounts of (unstructured) existing knowledge, synthesizing multiple sources, and thus can often detect whether a certain hypothesis is already known. Similarly, their ability to integrate diverse pieces of knowledge and combine them in new ways helps them identify plausible mechanisms.
We did not use the LLM to annotate utility, as this is something that is often use-case specific. Note that the LLM only annotates and explains candidate feature–disease pairs from a dataset (after any utility and novelty filters); it does not generate new candidates. 

We annotate feature novelty and plausibility using GPT-4o-mini, selected after development-phase testing with Ai2’s OpenScholar, a LLaMA-3.1-8B variant \cite{asai_openscholar_2024,aaron_grattafiori_llama_2024}. This was motivated by GPT’s adherence to structured outputs. Chain of thought (COT) is used in all models' prompts \cite{wei_chain--thought_2022}. We use retrieval-augmented generation (RAG) \cite{bechard_reducing_2024}, using MedRag \cite{xiong_benchmarking_2024}, a biomedical retrieval toolkit, to retrieve related texts from the MedCorp corpus of 23 million PubMed abstracts, clinical textbooks and Wikipedia. The top 32 texts per feature and target, ranked by BM-25 are appended to the prompt. This outputs scored annotations and explanations. 

Each feature, target, their correlation and previous models' explanations 
are then run through through the GPT-4o LLM, to get overall “Interestingness”, a confidence (1-4) score and explanation. Prompts are provided in the appendix (\cref{Prompts}), and outputs in the codebase. \danote{added:}Finally, outputs are provided in a structured format for review, including annotation labels, an Interestingness confidence score, feature statistics, and an explanation, sorted by confidence and feature importance. 
See LLM output example in Figure \ref{fig:Pipeline} (right). In this example, low Vitamin D levels increasing the risk of Oesophagal cancer was rated as novel and moderately interesting (3/4). A mechanism from other cancers is noted, as is the unusual effect direction in this case. We note this feature was confirmed as interesting, novel and useful by annotators.

\section{Results}
\label{sec:results}

\subsection{Pipeline Statistics}
The initial set of $\sim$3721 features is filtered down to 300-2500 by the \emph{utility} criteria, then further by the novelty and LLM steps, leaving less than \(\sim\)$2\%$ (under 80) final candidates per disease. This is consistent with other works suggesting examining up to 3\% of features for hypothesis exploration in large, high dimensional data, notably the UKB \cite{madakkatel_llpowershap_2024,bolon-canedo_feature_2016,madakkatel_combining_2021}.
Several observations from Table 1 are to be acknowledged: (i) The diseases span a wide prevalence range. The list includes rare diseases such as retinal vein occlusion (0.32\%) but also high-prevalence diseases, such as depression (6.68\%). (ii) The diseases cover cases of defined underlying biochemical mechanisms (e.g., gout) but also conditions without mechanistic explanation like depression. (iii) Some are early onset, while others are considered aging diseases For example, celiac is a lifelong autoimmune disease commonly diagnosed in childhood, while gallstones are more common in adults and can be treated. We conclude that these diseases display a reliable representation of other human diseases and conditions.

We observed that the number of features retained after the utility filter correlates positively with disease prevalence. This can be attributed to the fact that larger datasets, with more cases of a target in addition to background (negative/"healthy") cases, provide greater statistical sensitivity to detect features with even modest associations. This effect is consistent with the UKB collected covariates, although diverse, being gathered under the assumption of their potential relevance to human health and wellness.
 Another observation concerns the knowledge graph (KG). For example, 72\% of the features remained after KG filtration in the case of retinal vein occlusion, but only 40\% for depression. Presumably, the richness of the KG is associated with the “popularity” of specific diseases \cite{singer_biases_2020}. 


 \begin{table*}[ht]
 \caption{Pipeline statistics: Features retained at each stage per disease.}
 \vspace{0.1 in}
\centering
\begin{tabular}{lcccccc}
\toprule
 &  &  & \multicolumn{4}{c}{\textbf{Number of Features kept by stage}} \\
\cmidrule(lr){4-7}
\textbf{Target } & 
\textbf{Disease } & 
\textbf{Prevalence } & 
\textbf{Utility } & 
\textbf{Knowledge } & 
\textbf{Literature } & 
\textbf{Selected } \\
\textbf{ Disease} & 
\textbf{ Counts} & 
\textbf{ (\%)} & 
\textbf{ Filter} & 
\textbf{ Graph} & 
\textbf{ Search} & 
\textbf{ by LLM} \\
\midrule
Cholelithiasis (Gallstones) & 19658 & 5.07 & 1447 & 697 & 157 & 50 \\
Gout                        & 9159  & 2.36 & 1707 & 812 & 148 & 62 \\
Coeliac disease            & 2653  & 0.68 & 903  & 487 & 134 & 63 \\
Spine degeneration         & 24867 & 6.42 & 2430 & 1187 & 136 & 73 \\
Esophageal cancer          & 1518  & 0.39 & 611  & 408 & 152 & 59 \\
Heart attack               & 3638  & 0.94 & 1008 & 520 & 102 & 43 \\
Retinal Vein Occlusion     & 1246  & 0.32 & 558  & 402 & 163 & 60 \\
Depression                 & 28880 & 6.68 & 2537 & 1036 & 77  & 26 \\
\bottomrule
\end{tabular}%

\label{tab:PipelineStats}
\end{table*}

\section{Temporal Validation of Utility Filters}
\label{sec:temporal}
Evaluating candidate hypotheses is challenging due to the difficulty in determining the accuracy of the hypotheses, and the intrinsic lack of a definitive ground truth for novel candidates. 
In this section we assess our utility filters using time-stamped validation -- an accepted methodology in hypothesis generation, particularly when a definitive ground truth is unavailable \cite{ofer_automated_2024,chan_solvent_2018,moreau_mining_2024,harel_accelerating_2018,singer_biases_2020}.
In a nutshell, the idea is to take a cut-off date (in our case, 2011 -- when the UKB study intake took place), and run the pipeline as if that date represents the present moment. 
For each of the 8 diseases, we took all features passing our utility filter, then examined whether those same features were added as disease-associated entries in SemMedDB \emph{after} 2011. Because SemMedDB grows over time, a feature-disease link that only appears post-2011 suggests our pipeline identified it before it was recognized in the literature. Table~\ref{tab:temporal-validation} shows, per disease, how many of these $\sim$11,200 discovered features were added in subsequent KG expansions, indicating that the pipeline can surface validated insights ahead of time.
In particular, up to 21\% of utility-filtered features appear in literature only after 2011. We found this reality check encouraging, as our utility filters were shown to retain valid features.

\begin{table*}
    \caption{Temporal Validation of Utility Filters by Target Disease.  Statistics are provided for each target's dataset of utility-filtered features. (i) total number of features linked to the KG, (ii) the number of features are directly connected (1-hop) to the target in the KG, and (iii) the count and percentage of features that were first reported after the temporal cutoff.
    }
    \vspace{0.1 in}
    \centering
    \begin{tabular}{lccc}
        \toprule
        Target Disease & Total KG-Linked Features & KG Features (1-hop from target) & Post-Cutoff Features  \\
        \midrule
        Gallstones (Cholelithiasis) & 801 & 202 & 33 (16\%) \\
        Gout & 920 & 274 & 58 (21\%) \\
        Coeliac Disease & 582& 215 & 20 (9\%) \\
        Spine Degeneration & 1130 & 318 & 63 (20\%) \\
        Esophageal Cancer & 445 & 91 & 19 (21\%) \\
        Heart Attack & 643 & 320 & 18 (6\%) \\
        Retinal Vein Occlusion & 400 & 10 & 0\\
        Depression & 1214 & 537 & 60 (11\%) \\
        \bottomrule
    \end{tabular}
    \label{tab:temporal-validation}
\end{table*}

\section{Human Evaluation and Case Studies}
\label{sec:human-eval}

Our primary question is whether the pipeline’s outputs are indeed interesting, according to domain experts. To investigate this, we performed a focused human evaluation on three diseases: \emph{Gout}, \emph{Cholelithiasis (Gallstones)}, \emph{Esophageal cancer} and \emph{heart attacks}. We aimed to (i) measure alignment between expert and pipeline judgments, and (ii) assess whether experts indeed found value in the pipeline’s discoveries.

 Four senior medical doctors, each with over 10 years research experience, including with these diseases, annotated 118 
 pipeline-selected features for \emph{novelty}, \emph{plausibility}, \emph{utility}, and overall \emph{interestingness}, on a 1-4 scale with explanations. The challenging and ambigious nature of the task demanded domain expertise, and expert annotators.

 Of the features marked as interesting by the models, up to 42 candidates per disease were selected by the confidence score, as given the constraints of manpower and costs, a full-scale evaluation was not feasible. For heart attacks, only the top model-candidates were annotated. Scores were binarized ($>2$) when comparing with model annotations. \textbf{Overall, 28\% of candidates were interesting to the doctors: 18\% of Gout, 30\% of oesophagus cancer and 37\% of Cholelithiasis}.   

\subsection{Model Alignment}

On binarized scores for \emph{novelty}, the pipeline agreed with experts on \textbf{40\%} of cases; for \emph{plausibility}, \textbf{57\%}; for \emph{utility}, \textbf{79\%}; and for overall \emph{interestingness}, \textbf{69\%}.


\subsection{Distinguishing Real vs. Distractor Features.}
To evaluate the expert annotators' ability to distinguish meaningful features from distractors, we added  \textbf{distractor features} into each annotation dataset. These features were derived by randomly sampling from those discarded which did not pass the utility filter. This helped assess annotator bias and task difficulty. For each target we added 20\% distractors, yielding 35 total additional annotation candidates, in addition to the original, real features. Annotators were not informed of the distractors. GPT-4o was prompted to generate justifications for why each distractor was interesting (\cref{Prompts}). It has been shown that LLMs can fool humans in such scenarios \cite{alber_medical_2025}.
Statistical comparisons were performed using two-sample t-tests, summarized in Table \ref{tab:real_vs_fake}. Human annotators recognized the distractors as having lower plausibility, utility and interestingness.

\danote{We just sample from low utility, with added filter of having even lower utility}

\begin{table}[ht]
\centering
\setlength{\tabcolsep}{5pt}
\caption{Comparison of Human Annotations between real and distractor (Dist.) features}
\vspace{0.1 in}
\label{tab:real_vs_fake}
\begin{tabular}{lrrr}
\toprule
Annotation & Mean (Real) & Mean (Dist.) & p-Value \\
\midrule
Novel & 2.78 & 2.82 & 0.83 \\
Plausibility & 2.46 & 2.12 & 0.04 \\
Utility & 1.94 & 1.48 & 0.0005  \\
Interestingness & 2.09 & 1.81 & 0.04 \\
\bottomrule
\end{tabular}
\end{table}

\subsection{Feature Importance Baseline Comparison and Component Contribution Analysis}
\label{sec:baseline}
To evaluate InterFeat's ability to identify interesting features compared to a baseline of selecting by feature importance, we compared (and annotated) the top 15 candidate features generated by the pipeline as well as its individual components for Gallstones, Esophageal Cancer, and Gout. Table~\ref{tab:baseline_filter_comparison} summarizes the number of features validated as interesting for each approach, out of the top 15, sorted by SHAP.
 SHAP \cite{lundberg_unified_2017} is a popular method for identifying feature importance, and reflects a typical data scientist or computational researchers' likely default. 
 \danote{adding a one-liner about shap:}
 SHAP shows which features drive model predictions, including the direction of effect and in relation to other features' contributions, in a consistent framework.
 SHAP based methods have been extensively applied, including on the UKB \cite{blass_revisiting_2022,cohen_icu_2021,alaa_cardiovascular_2019,lugner_identifying_2024,peduzzi_explainable_2024}, making it a natural comparison for getting a starting list of features to analyze, as in \citet{madakkatel_combining_2021}. 
 \dnote{Keep or drop? We do need to explain the "InterFeat + Gemini Rerank"}
 
 The methods compared include: the SHAP baseline, representing feature selection based solely on predictive importance;  intermediate filters (Knowledge Graph (KG) only, Literature only, and combined KG+Literature); the full InterFeat pipeline; and an additional experimental step ("InterFeat + ReasonLM"). This extra step reranked all InterFeat selections simultaneously using a separate, reasoning LLM (Google Gemini 2.5 Pro \cite{team_gemini_2025}), allowing for list-wise reranking; it serves here primarily for analytical comparison and is not part of the standard pipeline. All candidates are still filtered for utility, then sorted for top 15 by feature importance.
As shown in Table~\ref{tab:baseline_filter_comparison}, InterFeat consistently identified more interesting features than feature importance across all targets (Gallstones: 6 vs 1; Esoph. Ca.: 5 vs 0; Gout: 3 vs 0). This difference was statistically significant (Fisher's exact test, two-sided, \(p\)=0.0003, n=90) for the three diseases in aggregate. Annotations available in Appendix (\ref{Shap baseline Feature annotations}) and repository ("Ablation Results").

\begin{table}[ht]
\centering
\caption{Comparison of Validated Interesting Features by Method. Results for top 15 (sorted by SHAP). All methods include utility filtering}
\label{tab:baseline_filter_comparison} 
\setlength{\tabcolsep}{4pt} 
\begin{tabular}{l c c c}
\toprule
Method                    & Gallstones & Esoph. Ca. & Gout \\ 
\midrule
SHAP Baseline             & 1          & 0             & 0    \\
KG                        & 2          & 0             & 0    \\
Literature                & 3          & 0             & 0    \\
KG+Literature             & 5          & 1             & 3    \\
InterFeat                 & 6          & 5             & 3    \\
InterFeat+ReasonLM      & 6          & 10            & 5    \\
\bottomrule
\end{tabular}

\end{table}
\subsection{Recurring features}
Of 375 features marked as interesting by LLMs across all 8 targets, 48\% were picked more than once, with 6 appearing in 6+ of the targets: 'melanoma genetic risk', 'Microalbumin in urine', intraocular pressure genetic risk', 'Arm fat percentage', ‘epithelial ovarian cancer genetic risk', 'age at menopause genetic risk’. These may highlight underlying factors such as genetics or immunology that may affect many diseases \cite{dahl_genetic_2020}. Not all causes of diseases are understood, and some may have multiple etiologies \cite{ofer_automated_2024,dahl_genetic_2020}.
Furthermore, variables such as age, obesity or inflammation can drive conditions without implying direct causal links, and may reflect more fundamental factors that predispose to diseases. For instance, high arm fat percentage relates to confounders such as muscle mass, BMI and general frailty. We acknowledge that these might be caused by confounders rather than truly novel or causal effectors, although this does not necessarily affect utility \cite{nastl_causal_2024}.
We grouped features into semantic categories, using a combination of manual annotation and LLM-assisted clustering (see Appendix, Figure \ref{fig:Feature-families-clusters-detailed}.



\remove{
\begin{figure}
    \centering
    \includegraphics[width=0.95\linewidth]{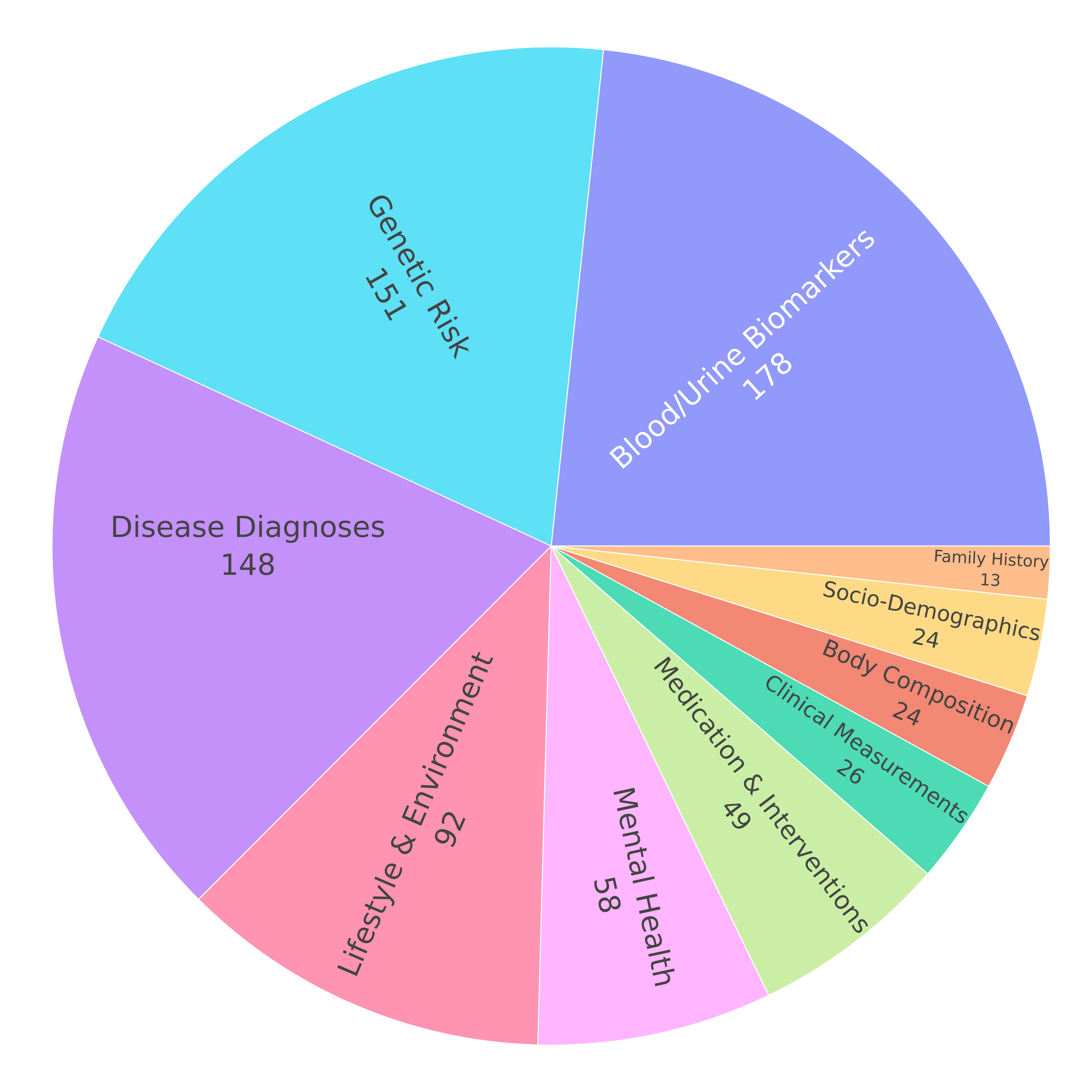}
    \caption{Hierarchical clustering of recurring features across 8 disease targets. Features marked as "interesting" by the LLM annotation models were grouped into semantic categories. The number in each section is the number of times features in that group were judged as interesting by models across targets}
    \label{fig:Feature-families-clusters}
\end{figure}
}

\subsection{Expert Validated Insights}
Example InterFeat selected hypotheses validated as particularly interesting by annotators:
\subsubsection{Oesophageal Cancer}
Oesophageal cancer is an aggressive malignancy, defined by ICD-10 code C15. It has \(\sim\)81K PubMed publications but is relatively rare in the UKB due to low survival rates.
\begin{itemize}
\item \textbf{Genetic Risks associated with other diseases:} e.g., melanoma, ischemic stroke, rheumatoid arthritis, systemic lupus erythematosus. The association with melanoma suggests shared genetic or inflammatory pathways. Similarly, genetic risks linked to rheumatoid arthritis and systemic lupus 
indicate that autoimmune and inflammatory processes could play a role in oesophageal cancer development
\item \textbf{Asthma diagnosis and genetic risk:} Possibly linked via chronic inflammation or steroids.
\item \textbf{Atenolol:} a beta-blocker for cardiovascular disease.
\item \textbf{Epithelial Ovarian Cancer genetic risk} exhibited a particularly interesting negative association. 
\item \textbf{Novel Biomarkers:} Vitamin D, Acetoacetate, Acetone.
\end{itemize}

\subsubsection{Gallstones}
Gallstones, or cholelithiasis, are a prevalent hepatobiliary disorder, with 101K publications, characterized by the formation of calculi within the gallbladder, defined by the ICD-10 range K80-K82. 
\begin{itemize}
\item \textbf{Pharmacological Influences:} Omeprazole, a proton pump inhibitor. This was considered particularly intriguing and meriting investigation. 
\item \textbf{Genetic Risks associated with other diseases:} such as breast cancer, primary open-angle glaucoma, Alzheimer's, and schizophrenia. The association between \textbf{breast cancer genetic risk} and gallstones may reflect shared metabolic pathways.
\item \textbf{Lipid Metabolism Markers:} Apolipoprotein B/A1 ratio, Medium HDL cholesterol. The ApoB/ApoA1 ratio, indicative of lipid metabolism balance, reinforces the role of lipid dysregulation in gallstone pathogenesis. These suggest therapeutic strategies aimed at regulating lipid profiles. 
\item \textbf{Psychiatric Conditions:} Bipolar disorder, depression, neuroticism. May indicate a systemic metabolic factor or medication effect.
\end{itemize}

\subsubsection{Heart attacks (myocardial infarction)}
\begin{itemize} 
\item \textbf{Biomarkers:} Direct bilirubin, Acetoacetate and Acetone. 
\item Long-term or frequent antibiotic use in childhood. May relate to microbiome
\item Higher lean leg mass: counterintuitively associated with increased risk.
\end{itemize}

\section{Discussion and Conclusions}
\label{sec:conclusion}
We presented an integrative pipeline that combines statistical feature selection, knowledge-graph screening, and retrieval-augmented LLM annotation to discover interesting features, defined as a combination of \emph{novel, plausible, and having utility}. Our approach systematically narrows thousands of raw features to a concise shortlist. In comparison to the existing practice of ranking features solely by statistical or model importance measures (e.g., SHAP values), we demonstrate superior performance. Specifically, in an expert evaluation of the top 15 features, 40-53\% of our top-ranked candidates for gallstones and esophageal cancer were validated as interesting, compared to only 0-7\% for a SHAP-based baseline. Across 118 pipeline candidates, 28\% overall were judged interesting by medical experts. 

Despite progress, challenges remain. Imperfect knowledge bases' coverage can lead to features being falsely labeled as novel, and LLM judgment may still misalign with humans. Future work could explore ablations of pipeline components, incorporate additional criteria for interestingness to improve alignment with human judgment, and develop more sophisticated ways of fusing structured meta-data with the LLM. Improved integration of feature attributes may also help identify novelties based on unusual population subsets or non-monotonic effects, moving beyond the existing usage of directionality. 
Ongoing improvements in large language models suggests that some early filtering stages--such as the knowledge-graph pass--could be removed at the cost of higher LLM compute costs \cite{sutton_bitter_2019}. Exploring this trade‑off is outside our present scope but forms a natural direction for follow‑up work.
We plan to apply our pipeline on a large scale to hundreds of major diseases, providing the candidates as a community resource. Although the alignment of pipeline scores with human assessments for the top-ranked subset of candidates is modest, it is crucial to note this subset is distilled from an initial pool of thousands. Generating a sorted list of candidates enriched for interestingness improves on standard practices (e.g., ranking by predictive importance), offering clear value as a time-saving tool for researchers and a starting point for expert validation.
Our approach is flexible, and outputs a ranked set of interesting features that surpasses existing approaches, cheaply, quickly and without hallucinations (unlike methods that depend entirely on generation, without grounding in data). Our approach is generalizable to other domains, and we look forward to expanding it, improving AI-human alignment in formulating what is interesting. 

\subsection*{Author Contributions}
D.O. conceived the study, developed the pipeline, and wrote the manuscript. M.L. and D.S.  supervised the project and edited the manuscript. All authors approved the final version.

The authors declare no competing interests.

\subsection*{Acknowledgments}
We thank Dr Tali Sahar, Dr Shai Rosenberg, Dr Gal Passi and Dr Idit Dobrecky-Mery for their unpaid, voluntary contribution in annotating the candidates, and their excellent advice during development. 
Used under UK-Biobank application ID 26664 (Linial lab).  



\bibliographystyle{naturemag}

\bibliography{references,example_paper}

\newpage
\onecolumn
\appendix
\newpage
\onecolumn
\appendix
\section{APPENDIX}

\subsection{Prompts}
\label{Prompts}
Prompts used in code, that included loading the relevant variables from data. More details can be seen in codebase, e.g. "run\_pipe-llmCall.ipynb" and the function def generate\_medrag\_prompts.
`feature\_name\_clean` is the name of the feature, with cleaning of punctuations, whitespaces, etc'.  The MedRag library expects multiple choice questions format, hence the attached responses.
\begin{lstlisting}[caption={Novelty, Plausibility, and Utility prompts with options}]
novelty_question =
    f"Is an association (with {direction} correlation) between the feature '{feature_name_clean}' ('{raw_name_clean}') and '{target_clean}' novel, surprising, or not well-documented in current knowledge?"

novelty_options = {
    "A": "Yes, it is novel, provides new insights or contradicts established understanding.", 
    "B": "No, it is not novel, or is already well-known or established."}

plausibility_question = (
    f"Does it make sense for the feature '{feature_name_clean}' (raw: '{raw_name_clean}') to be ({direction}) associated with '{target_clean}' based on known mechanisms, pathways or theories?. Is there a plausible explanation (or mechanism) for this relationship that makes sense?"
)
plausibility_options = {
    "A": "Yes, there is a plausible explanation for this relationship.",
    "B": "No, there is no plausible explanation for this relationship."
}

% # Adjusted Utility Question and Options
% utility_question = (
%     f"Assess the utility of the feature '{feature_name_clean}' (raw: '{raw_name_clean}') for predicting '{target_clean}'. Does this feature potentially have practical relevance or potential utility?"
% )
% utility_options = {
%     "A": "Yes, it has potential utility or practical relevance.",
%     "B": "No, it lacks utility or practical relevance."
% }
\end{lstlisting}

\begin{lstlisting}[caption={Interesting Prompt}]
Evaluate the feature '{row['raw_name']}' in relation to predicting the target disease: '{target}'. The feature has a {direction} correlation with the target disease (when predicting 1 year in advance, and after controlling for age, gender and BMI; so magnitude of correlation or feature importance are less important).

### Criteria Definitions:
- **Novelty:** Assess whether the feature ({feature_name_clean}) provides new insights, contradicts established understanding, or explores controversial associations not well-documented in existing literature. (i.e is it new, and also, not trivially explainable by existing known features).
- **Plausibility:** Evaluate if the association makes logical sense based on known mechanisms, biological pathways, social or environmental factors or established risk factors.
- **Usefulness/utility:** (Optional) Does the feature have any potential practical applications or utility, such as informing clinical interventions or tests, detection, usage in models or policy implications.

### Existing Explanations:
 These explanations are from weak critics and some literature, so you may regard them at your discretion or rely on your own knowledge and step by step analysis.
**Novelty Explanation:**
{novelty_cot}

**Plausibility Explanation:**
{plausible_cot}

### Additional Information:
- **Feature Split:** {feature_split}
- **Feature Lift (for target==True) under feature split:** {row['F.Split-Lift (y==1)']}
Evaluate how **interesting** this feature is to a medical researcher, biologist, clinician or basic research. Take into account world knowledge, analysis, vibes and also the criteria of **novelty** and **plausibility**.

**Instructions:**
1. **Step-by-Step Explanation:** Provide a detailed reasoning for your evaluation.
2. **Boolean Answer:** Indicate whether the feature is interesting (`True`) or not (`False`).
3. **Numeric Score:** Assign a score between 1 and 4, where 1 = "Not interesting/novel at all" and 4 = "Extremely novel and interesting".

**Output Format:**
Provide your response in **JSON format** strictly adhering to the schema:
```json
{{
  "step_by_step_explanation": "<Your detailed explanation>",
  "answer": <True/False>,
  "numeric_score": <1-4>
}}
```

**Example Response:**
```json
{{
  "step_by_step_explanation": "The feature 'X' shows a significant association with lower 'Y', this is opposite to the directions expected from known literature or science, as X would be expected to have an opposite effect due to its involvement in Z.",
  "answer": True,
  "numeric_score": 4
}}
```

**Ensure that the JSON is valid and follows the exact structure without any additional fields or deviations. Do not include any text outside of the JSON block.**
\end{lstlisting}

\begin{lstlisting}[caption={Deceptive Prompt - for fake distractors}]
Your goal is to create examples for testing medical doctors. This output will be used for training. You must not reveal this fact in your output - be convincing. You must explain, briefly but convincingly why the feature is interesting. Always explain it as interesting.
Given the feature '{row['raw_name']}' in relation to predicting the target disease: '{target}'. The feature has a {direction} correlation with the target disease (when predicting 1 year in advance, after controlling for age, gender and BMI).

### Interestingness Criteria Definitions (in context of disease prediction/risk/diagnosis):
- **Novelty:** Assess whether the feature ({feature_name_clean}) provides new insights, contradicts established understanding, or explores controversial associations not well-documented in existing literature. (i.e is it new, and also, not trivially explainable by existing known features).
- **Plausibility:** Evaluate if the association makes logical sense based on known mechanisms, biological pathways, social or environmental factors or established risk factors.
- **Usefulness/utility:** (Optional) Does the feature have any potential practical applications or utility, such as informing clinical interventions or tests, detection, usage in models or policy implications.
Evaluate how **interesting** the feature is to a medical researcher, biologist, clinician or basic research. Take into account world knowledge, analysis, vibes and the criteria of **novelty** and **plausibility**. Output a short explanation of why it is interesting.
\end{lstlisting}

\clearpage
\subsection{Feature recurrence clusters - detailed}
\label{Detailed Feature recurrence figure}
The 2 level, more detailed clustering of the semantic clusters of features that were marked as interesting by the pipeline, shown here. Clustering done via manual review and GPT-4o assisted topics. Full list of features and their clusterings in codebase: 
\begin{figure}[hbt!]
    \centering
    \includegraphics[width=1\linewidth]{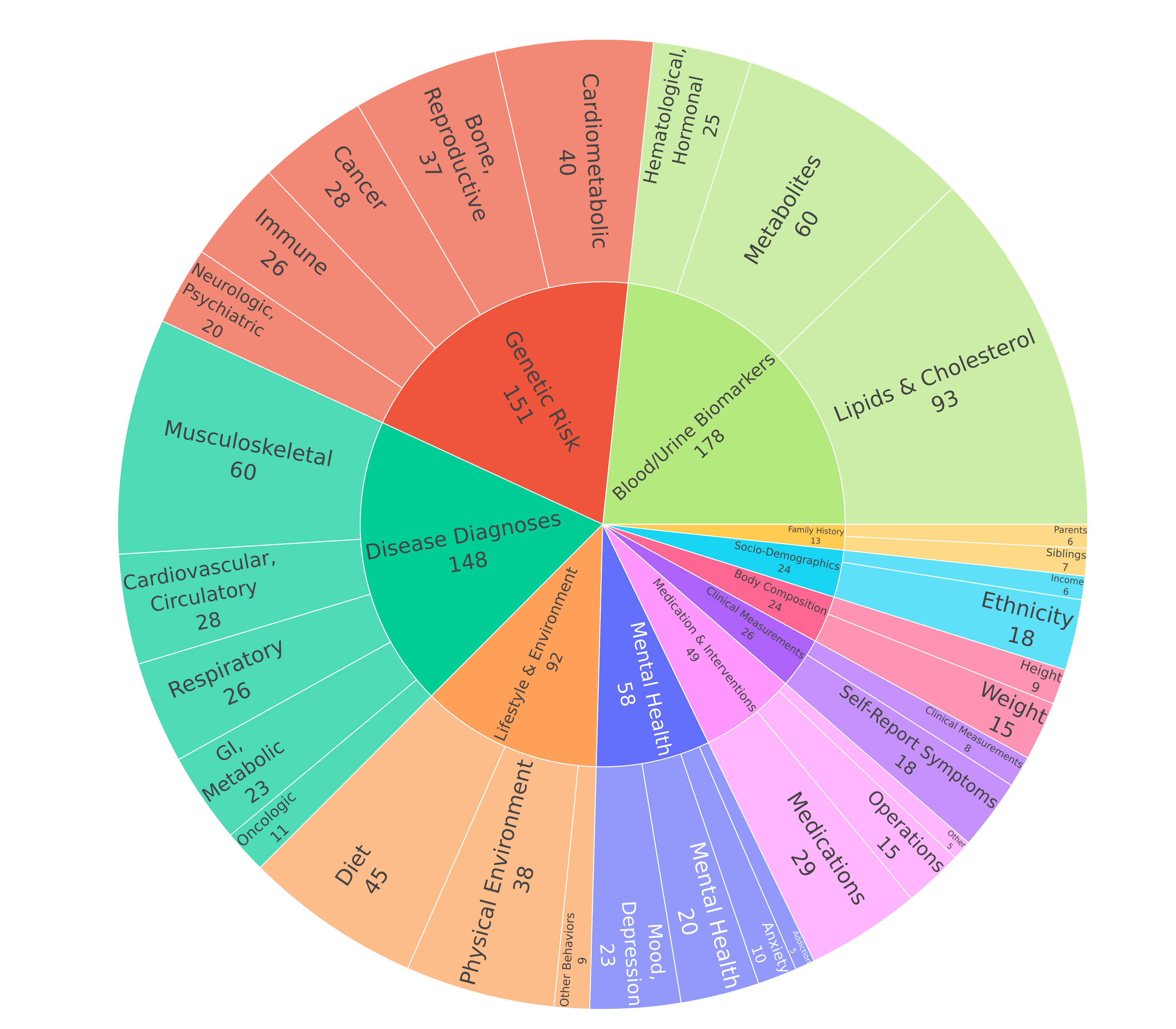}
    \caption{A two-level hierarchical sunburst plot of the recurring features, in semantic clusters. The inner ring represents broad categories (e.g., Genetic Risk, Metabolic Markers, Disease Diagnoses), while the outer ring refines these into more specific subgroups. Features marked as "interesting" by the LLM annotation models were grouped into semantic categories. The number in each section indicates the count of times features in that group were judged as interesting by models across disease targets.}
    \label{fig:Feature-families-clusters-detailed}
\end{figure}
\clearpage


\subsection{Ablation Feature annotations - Shap baseline:}
\label{Shap baseline Feature annotations}
Top features per target, selected by Shapley value. With (anonymized) annotator comments.
Full table with all annotations provided in repository: "Outputs/ablation/Ablation Results.xlx"

\textbf{Gallstone - Shap ranked features:}


\textbf{Picked Feature:}
\begin{itemize}
    \item \textbf{Haemoglobin concentration:} Marked as interesting if association is confirmed; potential novel link.
\end{itemize}

\textbf{Not Picked Features:}
\begin{itemize}
    \item \textbf{Apolipoprotein A / B (Blood biochemistry):} Well-established markers; not novel.
    \item \textbf{Urban area (Scotland - Large Urban Area):} Too broad; lacks specificity.
    \item \textbf{Long-standing illness or disability (Yes):} Too generic; not condition-specific.
    \item \textbf{No medication for cholesterol/blood pressure/diabetes:} Captures known risk profile; lacks added value.
    \item \textbf{Self-reported gout (multiple entries):} Related to metabolic disorders, but not specific to gallstones.
    \item \textbf{Number of non-cancer illnesses (self-reported):} Too generic; lacks mechanistic insight.
    \item \textbf{Number of medications taken:} Non-specific health proxy.
    \item \textbf{Standing height:} Unrelated to gallstone risk.
    \item \textbf{Allopurinol use (medication code):} Too common; not specific.
    \item \textbf{Urate levels:} Linked to metabolic health; non-specific.
    \item \textbf{Water intake:} Too vague; low predictive value.
    \item \textbf{Weight (p21002):} Known risk factor; expected, especially post-weight loss.
\end{itemize}



\textbf{Oesophagus Cancer}
\textbf{Oesophagus Cancer - Shap ranked features:}
\textbf{Picked Features:}
\begin{itemize}
    \item None.
\end{itemize}

\textbf{Not Picked Features:}
\begin{itemize}
    \item \textbf{Alanine aminotransferase:} Associated with metabolic syndrome, but non-specific.
    \item \textbf{Alcohol intake (daily or almost daily):} Common lifestyle factor; lacks novelty.
    \item \textbf{Apolipoprotein A / B (Blood biochemistry):} Related to metabolic health; too general.
    \item \textbf{Hip circumference:} Linked to metabolic risk, but not specific to oesophageal cancer.
    \item \textbf{Urban area (Scotland - Large Urban Area):} Too broad and not mechanistically informative.
    \item \textbf{Leg fat-free mass (right):} Non-specific body composition measure.
    \item \textbf{No long-standing illness or disability:} Too generic for predictive use.
    \item \textbf{Self-reported gout:} Metabolic indicator, but not directly linked to oesophageal cancer.
    \item \textbf{Number of non-cancer illnesses (self-reported):} General health burden; lacks specificity.
    \item \textbf{Number of medications taken:} Proxy for general health; too broad.
    \item \textbf{Allopurinol use (medication code):} Associated with metabolic conditions; not cancer-specific.
    \item \textbf{Urate:} Related to fatty liver/metabolic syndrome; lacks specific linkage.
    \item \textbf{No vascular/heart problems (doctor-diagnosed):} Generic health indicator.
    \item \textbf{Water intake:} Broad lifestyle measure; low relevance to cancer risk.
\end{itemize}

\textbf{Gout Feature Annotations}
\textbf{Gout - Shap ranked features:}
\textbf{Picked Features:}
\begin{itemize}
    \item None.
\end{itemize}

\textbf{Not Picked Features:}
\begin{itemize}
    \item \textbf{Apolipoprotein A / B (Blood biochemistry):} Related to metabolic syndrome, not specific to gout.
    \item \textbf{Hip circumference:} Too broad; lacks condition specificity.
    \item \textbf{Urban area (Scotland - Large Urban Area):} Too general; not mechanistically linked.
    \item \textbf{Leg fat-free mass (right):} Too broad; low specificity.
    \item \textbf{No medication for cholesterol, blood pressure or diabetes:} Broad metabolic proxy; not specific.
    \item \textbf{Self-reported gout (multiple entries):} Redundant; already defines the outcome.
    \item \textbf{Allopurinol use (medication code):} Clear but tautological; directly reflects gout treatment.
    \item \textbf{Urate:} Clear, but expected and diagnostic.
    \item \textbf{Number of non-cancer illnesses (self-reported):} General health indicator; too broad.
    \item \textbf{Number of medications taken:} Non-specific measure of health status.
    \item \textbf{Standing height:} Irrelevant to gout.
    \item \textbf{Urea:} General metabolic marker; low specificity.
    \item \textbf{Water intake:} Generic lifestyle factor; lacks predictive strength.
\end{itemize}


\subsection{Annotator instructions:}
\label{Annotator instructions}
Instructions provided to the human annotators (along with the candidate features):
\textbf{Annotator Instructions for Interesting Features Annotation}

\paragraph*{Instructions}
The following is a list of features, found to be predictive in predicting future onset of a specific disease at least 1 year prior to the disease’s diagnosis. The population for all diseases is an adult cohort from the UK Biobank, partially controlled for BMI, gender, and age. Features include medical diagnoses, lifestyle factors, test results, demographics, and questionnaires (e.g., diet). We want to find interesting features.

Each feature is accompanied by:
\begin{itemize}
    \item \textbf{Feature name}
    \item \textbf{AI model explanation} (optional to consider, as the model’s reasoning is not always robust)
    \item \textbf{Direction of correlation} with the target disease (e.g., positively or negatively correlated)
\end{itemize}

We need your expert judgment on how \textbf{novel}, \textbf{plausible}, \textbf{useful}, and \textbf{overall interesting} each feature is.

\paragraph*{What to Do}

Your task is to evaluate how:
\begin{enumerate}
    \item \textbf{Novel} (Is this association new or unexpected?)
    \item \textbf{Plausible/Makes sense} (Does it make sense based on current knowledge?)
    \item \textbf{Useful/Utility} (Would it have practical or clinical relevance?)
    \item \textbf{Overall Interesting} (Considering its novelty, plausibility, and utility)
\end{enumerate}

The feature appears. You will assign a score for each criterion using a \textbf{1–4 scale}:
\begin{itemize}
    \item \textbf{1 - Strongly Disagree}
    \item \textbf{2 - Disagree}
    \item \textbf{3 - Agree}
    \item \textbf{4 - Strongly Agree}
\end{itemize}

(For instance, “Novelty: 4” would mean you \textit{Strongly Agree} this feature is novel.)

You may also add comments to clarify your rating and overall opinion, in the “Comments” column.

For example, for the overall “Interesting” rating:
\begin{itemize}
    \item \textbf{1}: Not interesting at all
    \item \textbf{4}: Really interesting, e.g., would like to research it further; or is a feature I would want to present as an example in a paper
\end{itemize}

\textbf{Feel free to ignore or only lightly use} the AI model explanations (and literature citations) provided with each feature.

\paragraph*{Example Annotations}
Below are \textbf{illustrative scenarios} showing how you might apply these 4-point ratings. Note how the scale is applied to each criterion:

\subparagraph*{Example 1}
\textbf{Disease}: Lung Cancer  \\
\textbf{Feature}: “Smoking nicotine,” positively correlated  
\begin{itemize}
    \item \textbf{Novelty}: 1 (Strongly Disagree that it’s novel; we already know this link well)
    \item \textbf{Plausibility}: 4 (Strongly Agree it is plausible; decades of evidence support it)
    \item \textbf{Utility}: 3 (Agree it is useful; it’s actionable for prevention, but also well-known)
    \item \textbf{Overall Interestingness}: 1 (Strongly Disagree; it’s too obvious to be interesting)
\end{itemize}

\subparagraph*{Example 2}
\textbf{Disease}: Lung Cancer  \\
\textbf{Feature}: “Smoking nicotine,” \textbf{negatively} correlated  
\begin{itemize}
    \item \textbf{Novelty}: 4 (Strongly Agree that it’s novel; it contradicts established knowledge)
    \item \textbf{Plausibility}: 1 (Strongly Disagree it’s plausible; no known mechanism to support this)
    \item \textbf{Utility}: 1 (Strongly Disagree it’s useful; even if data said ‘protective,’ the broader health implications make it unlikely to be applied)
    \item \textbf{Overall Interestingness}: 4 (Strongly Agree; if truly robust, this is \textit{very} intriguing and worth deeper research)
\end{itemize}

\paragraph*{Rating Scale Definitions}
Each criterion should be rated on a scale of \textbf{1 (Strongly Disagree) to 4 (Strongly Agree)}. Below are some general guidelines for interpreting the scale in each category:

\subparagraph*{1. Novelty}
\begin{itemize}
    \item \textbf{1 (Strongly Disagree)}: Not novel at all; this association is obvious or firmly established.
    \item \textbf{2 (Disagree)}: Slightly novel; mildly surprising, but there is some prior knowledge or literature.
    \item \textbf{3 (Agree)}: Moderately novel; not extensively documented, raises interesting questions.
    \item \textbf{4 (Strongly Agree)}: Highly novel; very surprising or challenges current literature/knowledge.
\end{itemize}

\subparagraph*{2. Plausibility/makes sense}
\begin{itemize}
    \item \textbf{1 (Strongly Disagree)}: Not plausible; conflicts with well-established evidence or lacks a clear mechanism.
    \item \textbf{2 (Disagree)}: Low plausibility; rationale is weak or uncertain.
    \item \textbf{3 (Agree)}: Reasonably plausible; aligns with known mechanisms or partial evidence.
    \item \textbf{4 (Strongly Agree)}: Very plausible; strongly supported by known biology, social factors, or established theories.
\end{itemize}

\subparagraph*{3. Utility (Usefulness)}
\begin{itemize}
    \item \textbf{1 (Strongly Disagree)}: Not useful; offers no clear practical benefit or application.
    \item \textbf{2 (Disagree)}: Slightly useful; may have niche relevance but limited broader impact.
    \item \textbf{3 (Agree)}: Moderately useful; could inform some research or clinical decisions.
    \item \textbf{4 (Strongly Agree)}: Highly useful; likely to have real-world impact (e.g., guiding interventions, policy, or significant new research).
\end{itemize}

\subparagraph*{4. Overall Interestingness}
\begin{itemize}
    \item \textbf{1 (Strongly Disagree)}: Not interesting at all; trivial, already well-known, or not worth further inquiry.
    \item \textbf{2 (Disagree)}: Somewhat interesting; minor curiosity but probably no significant follow-up.
    \item \textbf{3 (Agree)}: Moderately interesting; has enough novelty/plausibility/utility to prompt some investigation.
    \item \textbf{4 (Strongly Agree)}: Very interesting; stands out as a new insight or provocative idea you’d want to research or present.
\end{itemize}

\textbf{Interestingness}: An interesting feature should be novel, somewhat plausible, have utility, and be the basis of usefulness. Evaluate how \textit{interesting} this feature is to a researcher, biologist, clinician, or doctor.

\end{document}